\documentclass[10pt,twocolumn]{article}
\usepackage{amsmath}
\usepackage{amsfonts}
\usepackage{amssymb}
\usepackage{graphicx}
\title{\textbf{Dipolar dissociation dynamics in electron collisions with carbon monoxide}}
\author{\textbf{Dipayan Chakraborty$^1$}, \textbf{Pamir Nag$^2$} and \textbf{Dhananjay Nandi$^3$}\\ Indian Institute of Science Education and Research Kolkata, Mohanpur 741246, India\\ \small{email:$^1$dc14rs010@iiserkol.ac.in $^2$pamir1118@iiserkol.ac.in, $^3$dhananjay@iiserkol.ac.in}}

\date{}
\begin{document}
\twocolumn[
  \begin{@twocolumnfalse}
    \maketitle
    \begin{abstract}
      Dipolar dissociation processes in the electron collisions with carbon monoxide have been studied using time of flight (TOF) mass spectroscopy in combination with the highly differential velocity slice imaging (VSI) technique. Probing ion-pair states both positive and/or negative ions may be detected. The ion yield curve of negative ions provides the threshold energy for the ion-pair production. On the other hand, the kinetic energy distributions and angular distributions of the fragment anion provide detailed dynamics of the dipolar dissociation process. Two ion-pair states have been identified based on angular distribution measurements using VSI technique. 
    \end{abstract}
  \end{@twocolumnfalse}
]
\section{Introduction}

Ion-pair states are generally superexcited states of molecules embedded in the ionization continuum. Such superexcited states can be accessed through photon or electron collisions with isolated molecules that in turn may dissociates into positive and negative ions. \cite{kuchi} Photoion pair formation has been studied quite extensively over the past few decades. \cite{rev:suits} Main emphasis was to determine threshold energy for the process more accurately employing threshold ion pair production spectroscopy (TIPPS).\cite{shiell, martin} However, detailed dynamics have been obtained in the recent times due to the availability of high resolution ion pair imaging spectroscopy (IPIS). \cite{rev:suits, Hao} On the other hand, in the electron collisions, the same ion pair states can be accessed and detailed dissociation dynamics can be obtained by probing fragment anion using high resolution velocity slice imaging (VSI) spectroscopy. 

VSI technique has been successfully applied to study negative ion formation due to dissociative electron attachment (DEA) process.\cite{cl:pamir, CO2:pamir, CO:pamir} The same technique has been extended here to study dissociation of the long-range ion-pair states of CO. Unlike DEA process, the anion formation due to dipolar dissociation does not proceed via resonant electron capture.\cite{dea:illenberger} In the latter case, the incident electron transfers some energy to the molecule and excite it to ion-pair states that eventually dissociates into cation and anion. The ion-pair formation is possible as long as the excitation energy is equal and more than the asymptotic ion-pair dissociation energy. The minimum energy position of the ion-pair state usually far away from the equilibrium position of the neutral molecule in the potential energy curve formalism. The ion-pair state may be accessed via direct excitation or indirectly through predissociation of an initially excited Rydberg state of the neutral molecule. The indirect mechanism is more commonly applied in the studies of photoion pair formation.\cite{rev:suits} The detection of ion pair provides information on the electronic structure of a molecule and the dissociation dynamics of its exited states. For electron collision studies, both direct and indirect mechanism may be applicable as discussed in the present article. In electron collision with CO the ion-pair states can give rise to momentum matched anion and cation products, either C$^+$ and O$^-$ or C$^-$ and O$^+$ channels:
\begin{equation}
\text{CO} + e^- \rightarrow \text{CO}^* + \text{e}^- \rightarrow \left\{ 
\begin{array}{c}
\text{C}^+ + \text{O}^- +\text{e}^-\\
                  \text{C}^- + \text{O}^+ +\text{e}^-
\end{array}
\right.
\end{equation}

The ion pair formation from CO was reported by Vaughan \cite{vaughan} and Lozier \cite{lozier} in the electron collision studies quite long ago. In the dipolar dissociation range, Lozier \cite{lozier} observed equal intensity of C$^{+}$ and O$^{-}$ formation with threshold energy of 20.9 $\pm$ 0.1 eV. Successive studies by several groups \cite{hagstrum1, dorman} also confirmed similar threshold energy. However, no kinetic energy and angular distribution data are available till now. In the current study, both C$^-$ and O$^-$ have been observed in the dipolar dissociation range. However, the count rate for C$^-$ is too low to perform any meaningful VSI study and is not reported in the present article. It is well accepted \cite{lozier, dorman} that the anion formation in the electron collision studies with reported primary electron energy range can only be possible through ion-pair states.   Here, it is assumed that the O$^-$ ions are always accompanied by C$^+$ ions but to verify the claim conclusively a coincidence measurement is absolutely necessary. In this article, we first outline the method and provide detailed studies of the dipolar dissociation dynamics in the electron collisions with carbon monoxide (CO) using VSI.

\section{Instrumentation}

The O$^-$ ions produced due to dipolar dissociation are studied using highly differential time sliced velocity map imaging technique. The current experimental setup is similar to the previous report of Nandi \emph{et al.} \cite{rsi:DN} with minor modifications as described by Nag and Nandi.\cite{MST:pamir} The same setup has been used to study the dissociative electron attachment to Cl$_2$, \cite{cl:pamir} CO$_2$ \cite{CO2:pamir} and CO \cite{CO:pamir} in the recent time. In brief, the experimental setup consists of an electron gun, a Faraday cup to measure electron current situated in the same axis, a needle of 1 mm diameter to produce effusive molecular beam and a time of flight (TOF) based velocity map imaging (VMI) spectrometer. The needle directed towards the detector is placed in the spectrometer axis and perpendicular to the electron beam axis.
The basic theme of the experiment is the effusive molecular beam interacts perpendicularly with the magnetically collimated pulsed electron beam. As a result, ions are formed in the interaction zone that are pulsed extracted into the spectrometer and detected by a two dimensional position sensitive detector. The electrons are produced by thermionic emission and the energy of the electrons is controlled by a programmable power supply. Typical energy resolution of the electron beam is about 0.8 eV. The pulse width of the electron beam is about 200 ns and the repetition rate is 10 kHz. After passing through the interaction region the electrons are collected using the Faraday cup that measures the time averaged electron beam current. The VMI spectrometer is a three field time of flight (TOF) type mass spectrometer capable to map all the ions with a given velocity vector to a point on the detector irrespective of their place of birth. The detector consists of three micro channel plates (MCP) with Z-stack configuration and three layers delay line hexanode.\cite{hex1} The TOF of the detected ions are determined from the back MCP signal \cite{MST:pamir} whereas the x and y positions of the ions are calculated from the three anode layers of the hexanode placed behind  the MCPs. The TOF (t) and (x, y) position of each detected ions are stored in list-mode format (LMF) using the CoboldPC software from RoentDek.  
The experiments are performed under ultra high vacuum condition with base pressure as low as 10$^{-9}$ mbar and 99.9\% pure commercially available CO gas. 

To obtain the ion yield curve a different set of data acquisition system has been used. Only the MCP signal is used for this purpose. The MCP signal is first amplified by a fast amplifier and then fed to a constant fraction discriminator (CFD). The output of CFD is fed to STOP of a nuclear instrumentation module (NIM) standard time to amplitude converter (TAC) and the START pulse is generated by a master pulse which is synchronized with the electron gun pulse. The time difference between this START and STOP is the TOF of the O$^-$ ion. The output of the TAC is connected to a multichannel analyzer (MCA, Ortec model ASPEC-927). Finally, it is communicated to a computer via USB 2.0 interface used for data acquisition. Our own LabVIEW based data acquisition system \cite{MST:pamir} is used to obtain the mass spectra and the ion yield curve.

When the electrons are collied with the molecule, `Newton Sphere' of ion is formed. One can obtain the angular distribution information from the projection of the `Newton spheres' onto a two-dimensional position sensitive detector. Ions with higher kinetic energy will fall onto the detector with bigger diameter. In the current experiment, a moderate pulsed extraction field is applied and negative ions are extracted from the source region of the spectrometer. The extraction pulse duration is 2 $\mu$s and applied 100 ns after the electron beam pulse. This delayed extraction provides sufficient time to expand the `Newton Sphere\,' so that we can obtain better time sliced images and also prevent the electrons from reaching the detector. The aim is to obtain the central slice of the Newton sphere containing the kinetic energy and angular distribution information of the detected ions. To obtain the central slice a suitable time window has been selected during offline analysis using CoboldPC software. These sliced images contain the ions ejected in the plane parallel to the detector and containing the electron beam axis. The typical full width at half maximum (FWHM) in TOF of the O$^{-}$ is about 500 ns and a 50 ns time window has been selected for slicing purpose. For low energy ions a thiner slice (25 ns) may be less erroneous. The electron energy calibration has been done using the resonant peaks of O$^{-}$/O$_{2}$ at 6.5 eV and the O$^-$/CO at 9.9 eV.\cite{ref:rapp} The calibration for the kinetic energy distribution measurements have been performed using the kinetic energy released by O$^-$/O$_2$ at 6.5 eV.\cite{o2:dn_cross} Further, this energy calibration has been checked by measuring the kinetic energy of O$^-$ ion produced by dissociative electron attachment to CO$_2$ \cite{co2:slaughter,CO2:pamir} at 8.2 eV.

\section{Results and Discussion}

\begin{figure}
\includegraphics[scale=.42]{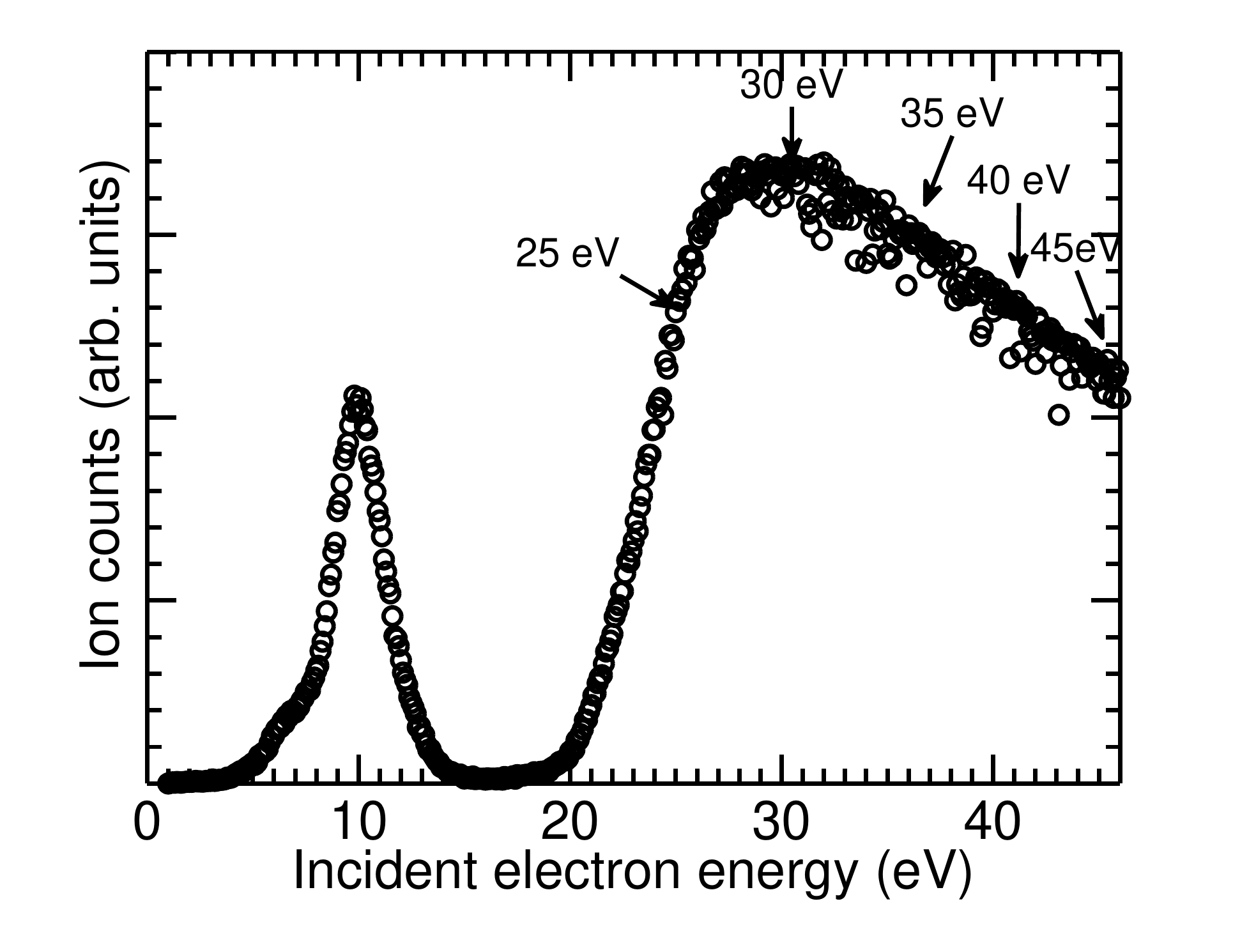}
\caption{\small{Ion yield curve of O$^-$ ion produced due to electron collision with gas phase CO molecule. The arrows indicate the energies at which the images are taken.}} \label{ion_yield}
\end{figure} 

Fig. \ref{ion_yield} shows the ion yield curve of the O$^-$ ions produced from CO due to 0-45 eV energy electron collisions. The ion yield curve is in good agreement with previous report.\cite{ref:rapp} A resonant peak at 9.9 eV due to dissociative electron attachment (DEA) is observed. The detailed DEA dynamics have been recently studied and reported elsewhere.\cite{CO:pamir} Increasing the electron energy revealed an access to the dipolar dissociation (DD) process that results the feature observed in the O$^{-}$ ion yield curve. The main focus of the present study is to understand the detailed dynamics occurring at the DD region, i.e., the process beyond 18 eV incident electron energy. In Fig.~\ref{ion_yield}, the arrows indicate the electron energies at which the VSIs are taken. In the following the threshold behaviour of the DD process seen in the ion yield is discussed within the limited energy resolution. The kinetic energy and angular distribution data extracted from VSI of the negative ions formed due to DD process are analysed thoroughly.

\begin{figure}
\centering
\includegraphics[scale=.42]{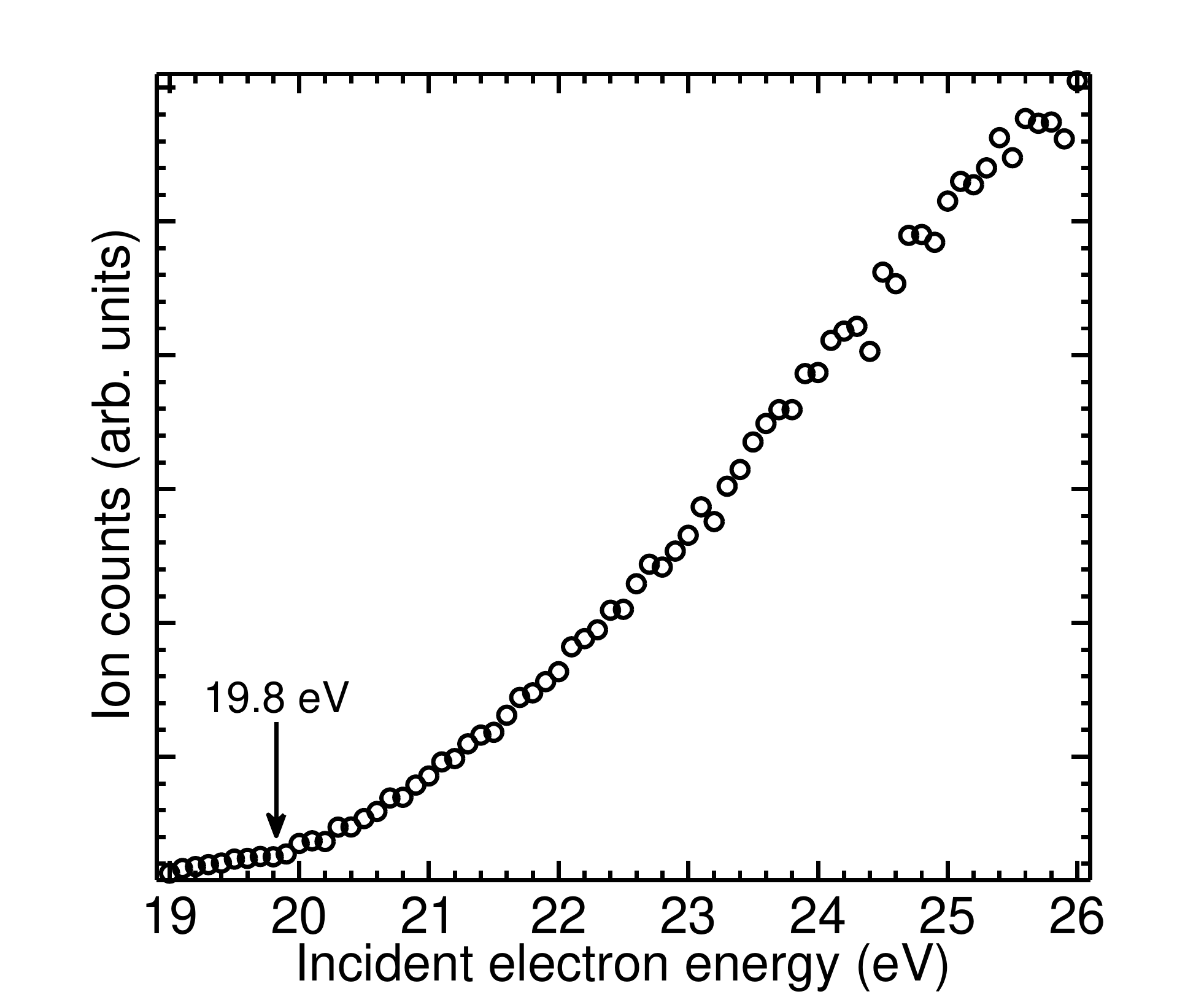}
\caption{\small{Ion yield for the O$^-$ ions produced in the dipolar dissociation range. The small circles represent the experimental data points . The threshold energy is shown by the arrow.}} \label{threshold2}
\end{figure}

Fig.~\ref{threshold2} shows the ion yield curve around the threshold of the DD region. Due to the finite energy resolution of the electron beam instead of being sharp, the curve gets smooth near the threshold value. From the experimental data the appearance energy of the anions is found to be near 19.8 eV. The appearance energy for the DD process can be calculated using the accepted values of the thermochemical parameters. \cite{web:nist}Using the conservation of energy, one can write the following expression for the DD process as,
\begin{equation}
V_e=(E_i+D-A+IP)+E_1 +E_2  \label{eq:threshold}
\end{equation} 
where $V_{e}$ is the amount of energy transfer from incident electron to molecule, $E_{i}$ is the energy associated with the possible excited states of the cation, $D$ is the bond dissociation energy, $IP$ is the ionization potential of carbon atom and the electron affinity of oxygen atom is $A$. The $E_1$ and $E_2$ are the kinetic energy associated with the C$^+$ and O$^-$ ions, respectively, at threshold both $E_1$ and $E_2$ are zero. Assuming both C$^+$ and O$^-$ ions are formed in ground state, \cite{locht, lozier, hagstrum1} i.e. $E_i=0$ the threshold can be calculated using the expression
\begin{equation}
E_{Th}=(D-A+IP)
\end{equation}
Using the values for thermochemical parameters \cite{web:nist} threshold energy for DD process of CO molecule can be obtain as $20.8$ eV. Within the experimental uncertainties the observed threshold value 19.8 eV matches well with the thermochemically obtained value. The minor deviation could be due to the contact potentials and finite resolution of the primary electron beam.

\begin{figure*}

\includegraphics[scale=.4]{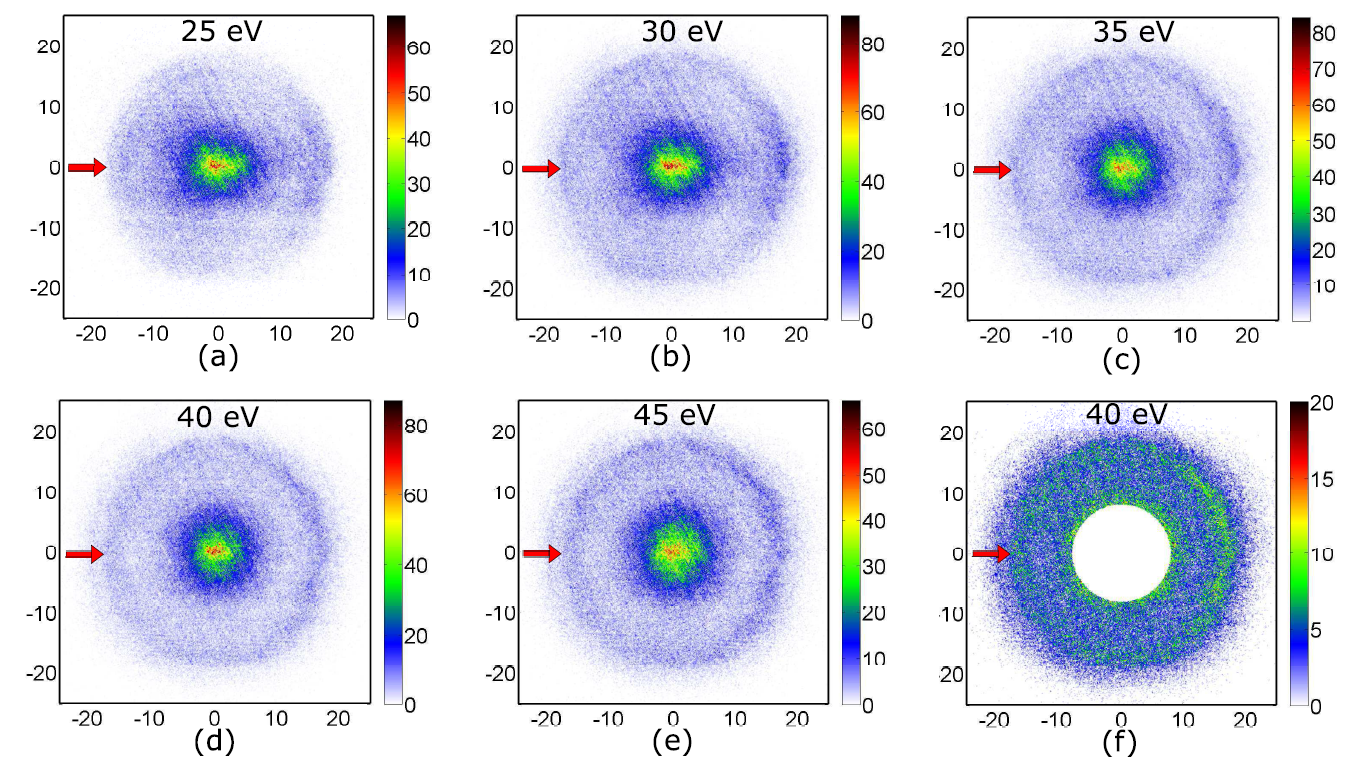}
\caption{\small{(a)-(e): Time sliced images taken with 50 ns time window of O$^-$ ion created due to the ion pair production at the indicated incident electron energies. (f) represents the same image as shown in (d) but without low energy part for better perspective. The arrows indicate the electron beam direction.}} \label{VSI_Im}
\end{figure*}

The time sliced velocity map images for different incident electron energies as indicated are shown in Fig. \ref{VSI_Im}. The kinematically complete information about the DD process can be obtain from these images. For the ion pair formation process, both anion and cation are formed. Images may be taken for either the cationic or anionic fragments as the conservation of linear momentum dictates that they both contain same information. In the present study only the anionic fragments are consider since it gives lower background and conclusive indication of the ion pair formation while cation may arise form other processes as well. Fig.~\ref{VSI_Im} (a)-(e) represent the velocity slice images (VSI) of the fragment anions taken at 25, 30, 35, 40 and 45 eV incident electron energies, respectively. Notice that all these images are taken with a constant time window of 50 ns width through the center of the respective Newton's spheres. Fig.~\ref{VSI_Im} (f) displays the same velocity slice image as Fig.~\ref{VSI_Im} (d) except the low kinetic energy part for better perspective. The incident electron beam direction is along the center of each image and from left to right as indicated by an arrow. Close inspection of each image shows a maximum intensity at the center and a ring pattern with larger diameter signifying the production of ions having two kinetic energy bands. The diameter of central pattern as well as annular pattern remain almost unchanged with increasing incident electron energy. These observations indicate two different mechanisms for the ion pair formation.       

\begin{figure}
\centering
\includegraphics[scale=.42]{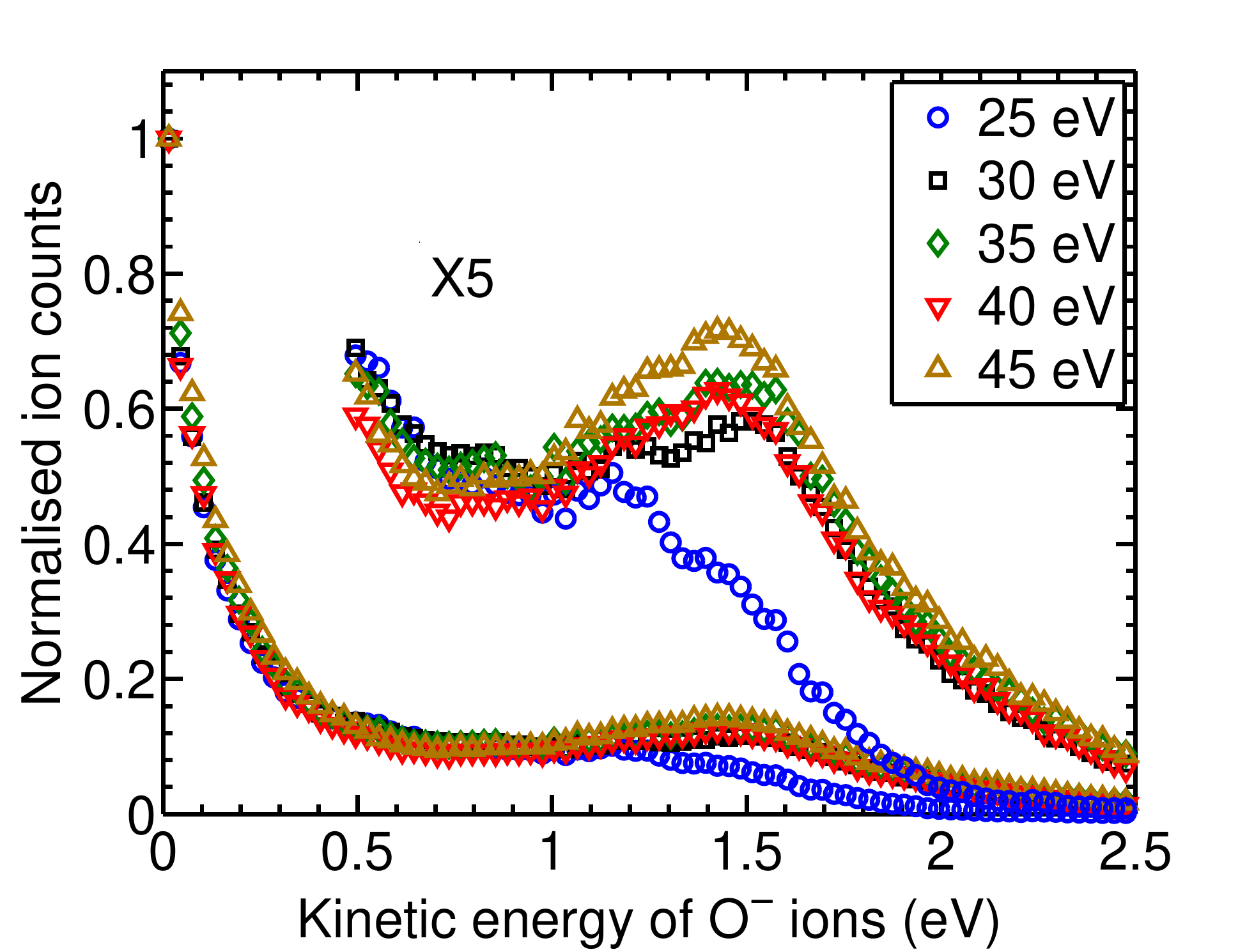}
\caption{\small{Kinetic energy distribution of the O$^-$ ions created due to dipolar dissociation process for five different incident electron energies.}} \label{KE_dist}
\end{figure}
 
The kinetic energy distributions of the O$^-$ ions have been extracted from the above sliced images and are shown in Fig.~\ref{KE_dist}. The distributions have been normalised near zero eV. One strong peak near zero eV followed by another broad band between 0.7 to 2.0 eV are observed. In order to obtain a better perspective of the second kinetic energy peak, the region 0.5 to 2.5 eV, in Fig.~\ref{KE_dist}, has been magnified by 5 times. Low and high kinetic energy bands arises in the kinetic energy distribution can be explained by the formation of indirect and direct ion pair process respectively. For the low kinetic energy bands the molecule first excites into a Rydberg state which crosses the ion pair state near the ion-pair dissociation limit. This results the predissociation of the Rydberg state via the ion pair state. In this case the dynamics of the ion pair dissociation is restrict by the degree of coupling between the initially excited Rydberg state and the ion pair state. The presence of low kinetic energy ions clearly indicate that the predissociation process occurs throughout the entire energy range. The higher kinetic energy band occurs due to the direct excitation to the ion pair state. The dynamics of the direct process is determined by the Franck-Condon factor. The initial increase of the kinetic energy with increasing electron energy is due to the access at different repulsive part of the ion pair state. One can calculate the appearance energy  for the direct excitation to the ion pair state by calculating the total kinetic energy release by the molecule. From Fig.~\ref{KE_dist} it can be observed that the higher kinetic energy band centred at 1.5 eV. From conservation of energy and momentum the kinetic energy of C$^+$ ion accompanying a O$^-$ ion of 1.5 eV can be found to be 2 eV. Using the values in expression (2) one can obtain that nearly 24.4 eV energy is transferred from the incident electron to the molecule. So in the Franck-Condon transition region the separation between the ground state of CO molecule and the ion pair state is around 25 eV. The truncated shape seen at 25 eV clearly indicates that such an ion pair resonant state enters into the Franck-Condon transition region around that energy.\
From Fig.~\ref{KE_dist} it can be observed that the kinetic energy of the ions remains unchanged for both lower energy as well as for higher energy with increasing incident electron energy. The possible explanation for this behaviour is as the incident electron energy increases, the available energy for the system is increases. Which can turn on some other excitation and ionisation process but the ion pair production proceeds through the population of the very same excited states throughout the energy range. So from these observations one can conclude that for incident electron energy near threshold  the ions are formed due to the indirect excitation process whereas, the direct excitation starts near 25 eV of incident electron energy.

\begin{figure}
\centering
\includegraphics[scale=.36]{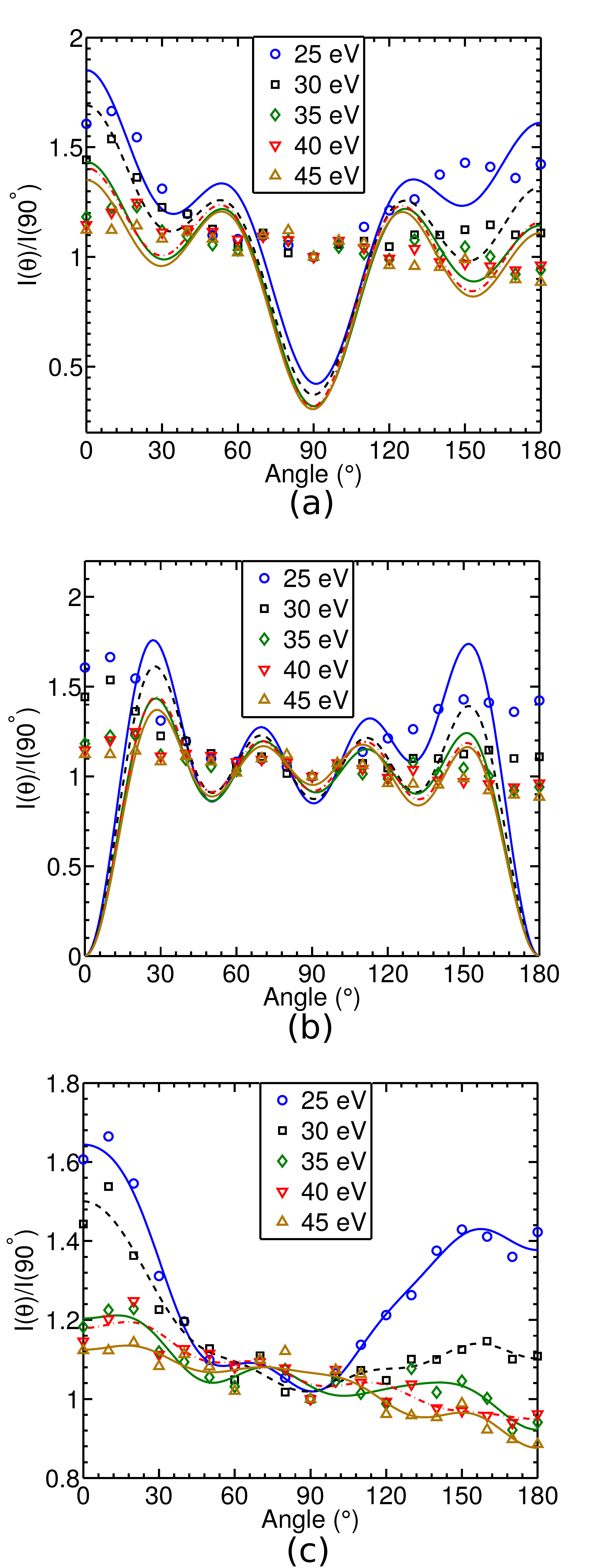}
\caption{\small{Angular distribution of the low kinetic energy O$^-$ ions created due to the ion pair formation process. Angular distribution data for all incident electron energies are fitted with (a) $\Sigma$ to $\Sigma$ transition, (b) $\Sigma$ to $\Pi$ transition, and (c) $\Sigma$ to $\Sigma + \Pi$ transition. Symbols represent the data points and lines are the fitted curve.}} \label{Ang-dist1}
\end{figure}

\begin{figure}
\centering
\includegraphics[scale=.36]{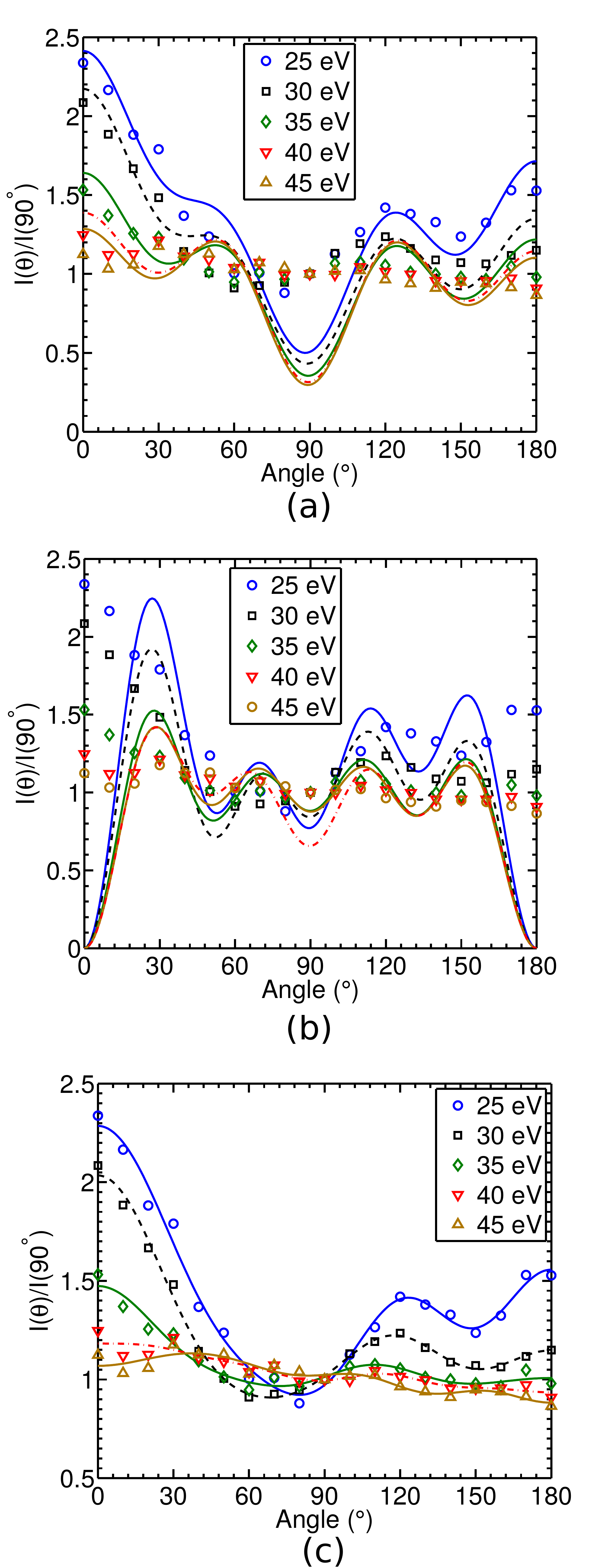}
\caption{\small{Angular distribution of the high kinetic energy O$^-$ ions created due to the ion pair formation process. Angular distribution data for all incident electron energies are fitted with (a) $\Sigma$ to $\Sigma$ transition, (b) $\Sigma$ to $\Pi$ transition, and (c) $\Sigma$ to $\Sigma + \Pi$ transition. Symbols represent the data points and lines are the fitted curve.}} \label{Ang-dist2}
\end{figure}

The angular distribution of fragment O$^-$ ions coming from the ion pair formation process have been analysed for both the kinetic energy bands. The observed angular distribution from VSIs taken at indicated electron energies and within the O$^{-}$ kinetic energy range of 0-0.4 eV and 0.7-2.3 eV are shown in Fig.~\ref{Ang-dist1} and Fig.~\ref{Ang-dist2}, respectively. A 25 ns and 50 ns thin time slices have been considered for the angular distribution analysis for low and high kinetic energy bands, respectively. In a flat slicing technique for the ions with higher kinetic energy only a fraction of the entire `Newton Sphere' is considered. But the entire `Newton Sphere' contributes in the sliced images of low kinetic energy ions.\cite{co2:moradmand2, inst:moradmand} In order to minimise this effect, thinner slices must be used for low kinetic energy ions. We used VSIs taken with 25 ns slice (not shown here) for the angular distribution of low energy ions. In the Figs.~\ref{Ang-dist1} and \ref{Ang-dist2}, symbols are the experimentally obtained data points and the solid curves are the fit-to-data using the model discussed below. All the data points have been normalised at 90$^{\circ}$. For low kinetic energy ions one dominant forward lobe and one backward lobe are seen whereas, for high kinetic energy ions one dominant forward lobe followed by two backward lobes are observed. However, the distribution becomes isotropic at the higher incident energies for both the cases. Similar results were observed previously by Van Brunt and Keiffer \cite{VanBrunt} and Nandi \emph{et al.} \cite{o2:polar_DN} in the study of oxygen molecule in the dipolar dissociation region. The reduction of anisotropy in angular distribution with increasing electron energy could be explained with the similar argument as given by Zare.\cite{zare} In the study of angular distribution from electron impact dissociation of H$_2^+$ ion, Zare concluded that the decreasing anisotropic nature is due to the $K$ (momentum transfer vector) dependence on $I(\theta)$ which is more near threshold. 

In order to obtain the symmetry of the ion-pair state(s) involved in the process, the angular distribution data have been fitted using the similar expression described by Van Brunt. \cite{VanBrunt1} According to Van Brunt \cite{VanBrunt1}, the experimental data can be fitted using the expression as,  
\begin{equation}
\small
I\left(\theta\right)=K^{-n}\left|\sum_{l=\left|\mu\right|}^{\infty}i^l\sqrt[]{\frac{\left(2l+1\right)\left(l-\mu\right)!}{\left(l+\mu\right)!}}j_l\left(\kappa\right)Y_{l,\mu}\left(\theta,\phi\right)\right|^2 
\label{fit-eq}
\end{equation}
This is similar with the equation derived by Zare.\cite{zare} Where $K$ is the momentum transfer vector between the incident and scattered electron, $\kappa$ denotes the product of the momentum transfer vector $K$ and the distance of closest approach between the impinging electron and the center of mass of the molecule, $j_l$'s are the spherical Bessel function, $Y_{l,\mu}$'s are the spherical harmonics and $\mu=\left|\Lambda_f-\Lambda_i\right|$, where $\Lambda_i$ and $\Lambda_f$ are the projection of the electronic axial orbital angular momentum along the molecular axis for the initial and final states, respectively. The $l$ is the angular momentum of the electron that is participating in the process. For hetero-nuclear diatomic molecules like the present case, $l\geq|\mu|$, whereas, for homo nuclear diatomic molecules, $l$ values are restricted to only even or odd depending upon whether the initial and final states are of same or opposite parity. The summation over $l$ takes care of the involvement of different partial waves. For a transition between two particular states for a given incident electron energy, the values of $K$ and $n$ are fixed and can be treated as parameters. Thus the angular distribution data due to the involvement of one or more than one final state(s) can be fitted using the expression as
\begin{equation}
\small
I\left(\theta\right)=\sum_{|\mu|}\left|\sum_{l=\left|\mu\right|}a_li^l\sqrt[]{\frac{\left(2l+1\right)\left(l-\mu\right)!}{\left(l+\mu\right)!}}j_l\left(\kappa\right)Y_{l,\mu}\left(\theta,\phi\right)e^{i\delta_l}\right|^2
\label{expression}
\end{equation}
Where $a_l$'s are energy dependent weight factors for the different partial waves, and $\delta_l$'s denote the phase differences between each partial wave responsible for the transition with respect to the lowest one. The summation over $\mu$ takes care of the involvement of more than one ion-pair states in the process. In present case, we considered upto four lowest partial waves for each of the state involved.

The solid curves in Fig. \ref{Ang-dist1} and \ref{Ang-dist2} represent the fit-to-data using the expression (\ref{expression}). The ground state of the neutral CO molecule is $^1\Sigma^+\left(\Lambda_i=0\right)$. The experimental angular distribution along with fits have been displayed in Fig. \ref{Ang-dist1} (a, b, c) by considering only $\Sigma$, only $\Pi$ and $\Sigma$ + $\Pi$ final state(s) and lowest four partial waves for the low kinetic energy ions. Similar data are displayed in Fig.~\ref{Ang-dist2} (a, b, c) for high energy ions. If we consider only $\Sigma$ final state transition, the angular distributions are well fitted in the forward and backward directions but underestimate in the perpendicular direction. Similarly, if we only consider a $\Pi$ final state transition, we underestimate in the forward and backward directions while angular distribution in the perpendicular direction fitted quite well. Finally, in order to obtain best fitted data, we need to consider both the $\Sigma$ and $\Pi$ final states transition. The above observations are applicable for both the low and high energy ions. However, we are unable to comment on whether same $\Sigma$ and $\Pi$ state are responsible for the two cases. The best fitted curve is shown in Fig.~\ref{Ang-dist1}(c) and Fig.~\ref{Ang-dist2}(c) for low and high energy ions, respectively. The values of different parameters used in the fit function are enlisted in Table \ref{table1} and Table \ref{table2} with $R^{2}$ values for low and high energy ions, respectively. In both the Tables, $a_0$, $a_1$, $a_2$, $a_3$ are the contribution of each partial wave for the transition to $\Sigma$ state, $b$'s are the contribution for transition to $\Pi$ state and $\kappa$'s are adjustable parameters. The observed  forward-backward asymmetry may be explained in the light of permanent dipole moment as described by Hall {\it et al.} \cite{hall} in the DEA to CO. A $\Sigma$ state contains a large negative dipole moment that favours backward peak in the angular distribution, whereas, a $\Pi$ state contains a positive dipole moment favouring forward peaking. The relative contribution of these two states may explain the observed forward-backward asymmetry.

\begin{center}
\begin{table*}
\small
\caption{Fitting parameters for the angular distribution of the O$^-$/CO ions arising from dipolar dissociation process with low kinetic energy and fitted with $\Sigma\rightarrow\Sigma+\Pi$ transition.}
\begin{tabular}{c c c c c c } \\ \hline
 & 25 eV & 30 eV & 35 eV & 40 eV & 45 eV\\ \hline
 Weighting ratio of & & & & & \\
  different partial waves & & & & & \\ 
$a_0:a_1:a_2:a_3:$ & 1.05:1.10:0.12:0.92: & 1.08:1.17:0.33:0.53: & 0.87:1.01:0.27:0.44: & 1.21:1.14:0.33:0.92: & 1.02:1.08:0.27:0.55: \\
$b_0:b_1:b_2:b_3$ & 1:1.36:1.80:0.09 & 1:1.14:1.18:0.06 & 1:1.17:0.60:0.23 & 1:1.20:0.63:0.02 & 1:0.91:0.70:0.03\\
 & & & & & \\
Phase difference ($\Sigma$) & & & & & \\
$\delta_{s-p}, \delta_{s-d}, \delta_{s-f}$ (rad) & 1.188, 1.269, 1.083 & 0.858, 1.233, 0.941 & 1.306, 2.072, 1.616 & 0.858, 1.755, 1.301 & 0.719, 1.456, 1.008\\
 & & & & & \\
Phase difference ($\Pi$) & & & & & \\
$\delta_{p-d}, \delta_{p-f}, \delta_{p-g}$ (rad) & 0.615, 0.707, 0.675 & 0.941, 0.498, 0.577 & 0.606, 0.266, 0.186 & 0.489, 0.148, 0.085 & 0.693, 0.163, 0.205\\
 & & & & & \\
 Parameter $\kappa_0, \kappa_1$ & 1.056, 1.524 & 1.091, 1.633 & 1.044, 1.31 & 1.308, 1.342 & 1.133, 1.411\\
 \hline
 $R^2$ value & 0.98 & 0.95 & 0.91 & 0.89 & 0.88\\ \hline
  \end{tabular} \label{table1}
 \end{table*}  
 \end{center}
  
\begin{center}
\begin{table*}
\small
\caption{Fitting parameters for the angular distribution of the O$^-$/CO ions arising from dipolar dissociation process with high kinetic energy and fitted with $\Sigma\rightarrow\Sigma+\Pi$ transition.}
\begin{tabular}{c c c c c c } \\ \hline
 & 25 eV & 30 eV & 35 eV & 40 eV & 45 eV\\ \hline
 Weighting ratio of & & & & & \\
  different partial waves & & & & & \\ 
$a_0:a_1:a_2:a_3:$ & 0.26:1.08:0.33:0.56: & 0:0.96:0.46:1.03: & 0.90:0.91:0.27:0.44: & 0.16:0.89:0.16:0.21: & 1.14:1.14:0.18:0.72: \\
$b_0:b_1:b_2:b_3$ & 1:0.82:0.98:0 & 1:1.15:0.48:0 & 1:0.90:0.58:0 & 1:0.36:0.99:0.04 & 1:0.44:1.14:0.04\\
 & & & & & \\
Phase difference ($\Sigma$) & & & & & \\
$\delta_{s-p}, \delta_{s-d}, \delta_{s-f}$ (rad) & 1.862, 1.442, 1.532 & 0.281, 0.276, 0.097 & 0.888, 1.344, 1.004 & 0.185, 0.376, 0.204 & 0.9364, 1.675, 1.275\\
 & & & & & \\
Phase difference ($\Pi$) & & & & & \\
$\delta_{p-d}, \delta_{p-f}, \delta_{p-g}$ (rad) & 1.082, 1.225, 0.074 & 0.825, 0.637, 0.567 & 1.064, 0.514, 2.072 & 2.287, 0.239, 0.271 & 1.853, 0.161, 0.296\\
 & & & & & \\
 Parameter $\kappa_0, \kappa_1$ & 1.083, 1.573 & 1.331, 1.255 & 1.096, 1.43 & 0.851, 1.646 & 1.169, 1.594\\
 \hline
 $R^2$ value & 0.99 & 0.99 & 0.96 & 0.87 & 0.88\\ \hline
  \end{tabular} \label{table2}
 \end{table*}  
 \end{center}

\section{Conclusion}
We have studied dipolar dissociation dynamics through ion-pair states of CO populated in electron collisions using velocity slice imaging, a well established method for dissociative electron attachment studies. The anion yield has been measured for threshold energy determination of ion-pair dissociation process. The threshold energy is in good agreement with previous reports. Velocity slice images have been taken at five different incident electron energies in the dipolar dissociation region. The kinetic energy and angular distributions have been extracted from the slice images. Low and high kinetic energy bands have been discussed using the indirect and direct ion pair formation process. A fixed kinetic energy release with increasing primary electron energy indicates that the molecule absorbs a fixed amount of energy from the incoming electron and that the rest of the energy is carried by the out going electrons. The truncated nature in the kinetic energy distribution at lower impact energy allow us to locate the position of the ion-pair states with respect to the ground state. Measured kinetic energy distributions clearly indicate that both direct and indirect ion-pair formation mechanism are responsible for the dipolar dissociation of CO. The angular distribution data strongly suggest the involvement of two ion-pair states in the studied electron energy range. The symmetry of these observed states are $\Sigma$ and $\Pi$ for both direct and indirect ion pair formation. We cannot conclude whether the same ion-pair states are involved or not in the direct and indirect ion pair formation. The peaking intensity in the forward and backward directions is most likely due to the $\Sigma$ final state transition. Whereas small peaks may arise from $\Pi$ final state transitions.  The theoretical calculations are badly needed to understand the detailed dynamics for the ion pair formation process.

\section{Acknowledgements}
D. N. gratefully acknowledges the partial financial support from ``Indian National Science Academy" for the development of VSI spectrometer under INSA Young Scientist project ``SP/YSP/80/2013/734".

\bibliographystyle{unsrt}

\begin{thebibliography}{10}

\bibitem{kuchi}
K.~Kuchitsu.
\newblock {\em Dynamics of excited molecules}.
\newblock Studies in physical and theoretical chemistry. Elsevier, 1994.

\bibitem{rev:suits}
Arthur~G. Suits and Jhon~W. Hepburn.
\newblock Ion pair dissociation: Spectroscopy and dynamics.
\newblock {\em Annu. Rev. Phys. Chem.}, 57:431, 2006.

\bibitem{shiell}
X.~K.~Hu R.~C.~Shiell, Q.~J. Hu, and J.~W. Hepburn.
\newblock A determination of the bond dissociation energy (d0(h−sh)): 
  threshold ion-pair production spectroscopy (tipps) of a triatomic molecule.
\newblock {\em J. Phys. Chem. A}, 104(19):4339--4342, 2000.

\bibitem{martin}
James~DD Martin and JW~Hepburn.
\newblock Electric field induced dissociation of molecules in rydberg-like
  highly vibrationally excited ion-pair states.
\newblock {\em Phys. Rev. Lett.}, 79(17):3154, 1997.

\bibitem{Hao}
Chang~Zhou Yusong~Hao and Yuxiang Mo.
\newblock Velocity map imaging study of the o2 ion-pair production at 17.499
  ev: Simultaneous parallel and perpendicular transitions.
\newblock {\em J. Phys. Chem. A}, 109:5832, 2005.

\bibitem{cl:pamir}
Pamir Nag and Dhananjay Nandi.
\newblock Identification of overlapping resonances in dissociative electron
  attachment to chlorine molecules.
\newblock {\em Phys. Rev. A}, 93:012701, 2016.

\bibitem{CO2:pamir}
Pamir Nag and Dhananjay Nandi.
\newblock Dissociation dynamics in the dissociative electron attachment to
  carbon dioxide.
\newblock {\em Phys. Rev. A}, 91:052705, 2015.

\bibitem{CO:pamir}
Pamir Nag and Dhananjay Nandi.
\newblock Fragmentation dynamics in dissociative electron attachment to co
  probed by velocity slice imaging.
\newblock {\em Phys. Chem. Chem. Phys.}, 17:7130--7137, 2015.

\bibitem{dea:illenberger}
E.~Illenberger and J.~Momigny.
\newblock {\em Gaseous Molecular Ions: An Introduction to Elementary Processes
  Induced by Ionization}.
\newblock Topics in Physical Chemistry. Steinkopff, 2013.

\bibitem{vaughan}
Alfred~L. Vaughan.
\newblock {\em Phys. Rev.}, 38:1687--1695, Nov 1931.

\bibitem{lozier}
W.~Wallace Lozier.
\newblock The heat of dissociation of co and the electron affinity of o.
\newblock {\em Phys. Rev.}, 46:268, August 1934.

\bibitem{hagstrum1}
Homer~D. Hagstrum and John~T. Tate.
\newblock Ionization and dissociation of diatomic molecules by electron impact.
\newblock {\em Phys. Rev.}, 59:354, February 1941.

\bibitem{dorman}
F.~H. Dorman, J.~D. Morrison, and A.~J.~C. Nicholson.
\newblock Threshold law for the probability of excitation by electron impact.
\newblock {\em J. Chem. Phys.}, 32:378, February 1960.

\bibitem{rsi:DN}
Dhananjay Nandi, Vaibhav~S. Prabhudesai, E.~Krishnakumar, and A.~Chatterjee.
\newblock Velocity slice imaging for dissociative electron attachment.
\newblock {\em Rev. Sci. Instrum.}, 76(5):053107, 2005.

\bibitem{MST:pamir}
Pamir Nag and Dhananjay Nandi.
\newblock Complete data acquisition and analysis system for low energy electron
  molecule collision studies.
\newblock {\em Meas. Sci. Technol.}, 26:095007, 2015.

\bibitem{hex1}
O.~Jagutzki, A.~Cerezo, A.~Czasch, R.~{D\"{o}rner}, M.~Hattas, Min Huang,
  V.~Mergel, U.~Spillmann, K.~Ullmann-Pfleger, T.~Weber,
  H.~Schmidt-{B\"{o}cking}, and G.D.W. Smith.
\newblock Multiple hit readout of a microchannel plate detector with a
  three-layer delay-line anode.
\newblock {\em IEEE Trans. Nucl. Sci.}, 49(5):2477--2483, Oct 2002.

\bibitem{ref:rapp}
Donald Rapp and Donald~D. Briglia.
\newblock Total cross sections for ionization and attachment in gases by
  electron impact. ii. negative-ion formation.
\newblock {\em J. Chem. Phys.}, 43(5):1480--1489, 1965.

\bibitem{o2:dn_cross}
Dhananjay Nandi and E.~Krishnakumar.
\newblock Dissociative electron attachment to polyatomic molecules: Ion kinetic
  energy measurements.
\newblock {\em Int. J. Mass Spectrom.}, 289(1):39 -- 46, 2010.

\bibitem{co2:slaughter}
D~S Slaughter, H~Adaniya, T~N Rescigno, D~J Haxton, A~E Orel, C~W McCurdy, and
  A~Belkacem.
\newblock Dissociative electron attachment to carbon dioxide via the 8.2 ev
  feshbach resonance.
\newblock {\em J. Phys. B: At. Mol. Opt. Phys.}, 44(20):205203, 2011.

\bibitem{web:nist}
{NIST}.
\newblock http://www.nist.gov/pml/data/handbook/index.cfm.

\bibitem{locht}
J.~Momigny R.~Locht.
\newblock Mass spectrometric study of ion-pair processes in diatomic molecules:
  H2, co, no and o2.
\newblock {\em Int. J. Mass Spectrom. Ion Phys.}, 7:121, August 1971.

\bibitem{co2:moradmand2}
A.~Moradmand, D.~S. Slaughter, A.~L. Landers, and M.~Fogle.
\newblock Dissociative-electron-attachment dynamics near the 8-ev feshbach
  resonance of co${}_{2}$.
\newblock {\em Phys. Rev. A}, 88:022711, Aug 2013.

\bibitem{inst:moradmand}
A.~Moradmand, J.~B. Williams, A.~L. Landers, and M.~Fogle.
\newblock Momentum-imaging apparatus for the study of dissociative electron
  attachment dynamics.
\newblock {\em Rev. Sci. Instrum.}, 84(3):033104, 2013.

\bibitem{VanBrunt}
R.~J. Van~Brunt and L~.J. Kieffer.
\newblock Electron energy dependence of the energy and angular distributions of
  o− from dissociative ion pair formation in o2.
\newblock {\em J. Chem. Phys.}, 60(7):3057--3063, 1974.

\bibitem{o2:polar_DN}
D.~Nandi, V.S. Prabhudesai, and E.~Krishnakumar.
\newblock Velocity map imaging for low-energy electron–molecule collisions.
\newblock {\em Radiat. Phys. Chem.}, 75(12):2151 -- 2158, 2006.

\bibitem{zare}
Richard~N. Zare.
\newblock Dissociation of h2+ by electron impact: Calculated angular
  distribution.
\newblock {\em J. Chem. Phys.}, 47(1):204--215, 1967.

\bibitem{VanBrunt1}
Richard~J. Van~Brunt.
\newblock Breakdown of the dipole‐born approximation for predicting angular
  distributions of dissociation fragments.
\newblock {\em J. Chem. Phys.}, 60(8):3064--3070, 1974.

\bibitem{hall}
R.~I. Hall, I.~\ifmmode \check{C}\else \v{C}\fi{}ade\ifmmode~\check{z}\else
  \v{z}\fi{}, C.~Schermann, and M.~Tronc.
\newblock Differential cross sections for dissociative attachment in co.
\newblock {\em Phys. Rev. A}, 15:599--610, Feb 1977.

\end{thebibliography}

\end{document}